\documentclass[a4paper,12pt,oneside]{article}
\usepackage{amsmath}
\usepackage{latexsym}
\usepackage[english]{babel}
\usepackage{graphics}
\usepackage{epsfig}
\usepackage{indentfirst}
\usepackage{cite}

\title{Photon Emission from Charged Particles Moving in Undulators Placed
in a Photonic Crystal}

\author{V.G. Baryshevsky, A.A. Gurinovich}

\date{}
\begin{document}
\maketitle
\begin{abstract}
The law of  photon radiation from electron beams passing through a
photonic crystal in the presence of an undulator is considered.
The spectral-angular distribution and the time evolution of
radiation is found.
It is shown that the characteristics of the radiation intensity in
time demonstrate oscillations due to the interference between
undulator and parametric (quasi-Cherenkov) radiations.
\end{abstract}

\section{Introduction}


It was shown in \cite{VG1971,VGID1972} that when a relativistic
particle moves in a crystal in which an electromagnetic wave
propagates, the radiation spectrum of the particle, oscillating in
the wave, changes significantly due to the virtual photon
diffraction, occurring in the process of emission in the crystal.

Moreover, even in the X-ray range, in which the refractive index
in a crystal is less than unity {${n(\omega)< 1}$}, under
diffraction conditions ${n(\omega)}$ can appear to be greater than
unity,
and so the induced Vavilov-Cherenkov effect becomes possible.

A relativistic oscillator can also be formed when a charged
particle is channeled through a crystal, causing channeling
radiation to appear \cite{VG+Dub1976,VG+Dub1977,Kumahov}
, or when a particle moves in a crystal undulator
\cite{VG+Grub+Dub:Ph.Lett,Vorobiev,Maisheev1,Maisheev2,VG+Tikh2013}.

At present, radiation from a relativistic oscillator in a crystal
in presence of photon diffraction is called the diffraction
radiation of a relativistic oscillator (DRO) or, in the case of
channeled particles,  the diffracted channeling radiation (DCR).
A thorough theoretical consideration of DRO and DCR was given in
\cite{VG+Dub:J.Phys.C,VG+Grub+Dub1,VG+Grub+Dub2,Nitta,Nitta2,Pivovarov}.

The spectral-angular distributions of radiation from channeled
particles were determined for Laue and Bragg geometries
\cite{VG+Grub+Dub1,VG+Grub+Dub2} (for more details see also
\cite{VG_NO2012}) and for the case of particle motion in a crystal
undulator in Laue geometry \cite{VG+Dub1991}.
In \cite{time1,time2} we considered the time evolution of the
radiation process in crystals (natural or photonic).
It was shown that  strong frequency dispersion of waves in
crystals can lead to a decrease in  group velocity of radiation in
crystals (the wave packet can slow down), and the situation can
arise when the radiation from the crystal continues even after the
particle has escaped from it. In \cite{time1,time2} a detailed
analysis was carried for the case of parametric (quasi-Cherenkov)
radiation.

The present paper studies the law of  photon radiation from
photonic crystals in the presence of an undulator. It also
discusses the law of time evolution  of DCR and radiation
generated through particle motion inside a crystal undulator.
The spectral-angular distribution of radiation when particles travel in the undulator formed in a photonic crystal  is found.
The formulas are derived that describe the laws of time evolution of radiation from a traveling particle.
%
It is shown that the characteristics of the radiation intensity in
time demonstrate oscillations due to the interference between
undulator and parametric (quasi-Cherenkov) radiations.
%

\section{Time dependence of the intensity of radiation from a particle moving in a crystal (natural or photonic)}


According to \cite{Landau7}, the radiation intensity $dI(t)$ in the element of solid angle
$d \Omega$ is defined as the amount of energy passing in unit time
 through the element $dS=r^2 d
\Omega$ of the sphere of radius $r$ that is much larger than the size of the radiation source and
has the origin of coordinates inside the radiation source.

The radiation intensity $dI(t)$ can be found if we know the
expression for the electric (magnetic) component of the
electromagnetic wave
$\vec{E}(\vec{r},t)$ ($\vec{H}(\vec{r},t)$),
produced by the radiation source \cite{Landau7}:
\begin{equation}
\label{1.1} dI(t)=\frac{c}{4\pi}|\vec{E}(\vec{r},t)|^2r^2 d
\Omega,
\end{equation}
where  $r$ is the distance from the observation point.

In the case we are considering here, the field  $\vec{E}(\vec{r},t)$ appears as a result of
interaction of a relativistic particle and a crystal.

The field $\vec{E}(\vec{r},t)$ can be expanded in a  Fourier
series
\begin{equation}
\label{1.2}
\vec{E}(\vec{r},t)=\frac{1}{2\pi}\int\vec{E}(\vec{r},\omega)e^{-i\omega
t}d\omega.
\end{equation}

According to the results obtained in
\cite{PXRbook,VG_NO2012,time2}, at large distance from the
crystal, the Fourier component $\vec E(\vec r,\omega)$ can be
written as
\begin{equation}
\label{1.3}
\vec{E}(\vec{r},t)=\frac{e^{ikr}}{r}\frac{i\omega}{c^2}\sum_{s}e_i^s
\int\vec{E}_{\vec{k}}^{(-)s^*}(\vec{r}^{\prime}\omega)\vec{j}(\vec{r}^{\prime},\omega)
d^3 r^{\prime},
\end{equation}
where  $i=1,2,3$ ($x$, $y$, $z$, respectively); $e^s_i$ is the
$i$-the component of the polarization vector, $s=1,2$;
\begin{equation} \label{1.4} \vec{j}(\vec{r},
\omega)=\int \vec{j}(\vec{r}, t)e^{i\omega t}dt,
\end{equation}
$\vec{j}(\vec{r}, t)= Q\vec{v}(t)\delta(\vec{r}-\vec{r}(t))$ is
the current density for a particle of charge  $Q$ and
 $\vec{r}(t)$ is the particle coordinate at  time $t$;
$\vec{E}^{(-)s}_{\vec{k}}$ is the solution of Maxwell equations
that describes scattering by a crystal of the plane wave having
the wave vector $\vec{k}=k\frac{\vec{r}}{r}$ and the wave number
$k=\frac{\omega}{c}$. At large distance from the crystal, the
solution $\vec{E}_{\vec{k}}^{(-)s}$ has a form of a superposition
of a plane and a converging spherical wave
\cite{VG_NO2012,PXRbook}.

The explicit expression for $\vec{E}^{(-)s}$ that describes the
diffraction of the  electromagnetic wave in a crystal for the Laue
and Bragg cases is given in \cite{VG_NO2012,PXRbook,VG1997}.

Let us discuss the following expression for the  the amplitude
$A(\omega)$ of the radiated wave in more depth
\begin{equation}
\label{1.5} A_{\vec{k}}^s(\omega)=\frac{i\omega}{c^2}\int
\vec{E}_{\vec{k}}^{(-)s^*}(\vec{r}^{\prime},\omega)\vec{j}(\vec{r}^{\prime},\omega)d^3
r^{\prime}.
\end{equation}
Using (\ref{1.4}) and  (\ref{1.5}), we can write
\begin{eqnarray}
\label{1.61}
A_{\vec{k}}^s(\omega)&=&\frac{i\omega}{c^2}\int \vec{E}_{\vec{k}}^{(-)s^*}(\vec{r}^{\prime},\omega)Q\vec{v}(t)
\delta(\vec{r}^{\prime}-\vec{r}(t))e^{i\omega t}dt d^3 r^{\prime}\nonumber\\
&=&\frac{i\omega
Q}{c^2}\int\vec{E}_{\vec{k}}^{(-)s^*}(\vec{r}(t),\omega)\vec{v}(t)e^{i\omega
t} dt
\end{eqnarray}

Let us recall here that
$\vec{E}_{\vec{k}}^{(-)s^*}=\vec{E}_{-\vec{k}}^{(+)s}$, where the
field $\vec{E}_{-\vec{k}}^{(+)s}$ is the solution of Maxwell
equations that describes scattering by a photonic crystal of a
plane wave, having the wave vector $(-\vec{k})$, and whose
asymptotic form at infinity is a superposition of a plane and a
diverging spherical wave.
In view of (\ref{1.61}), the radiation amplitude is determined by the field
$\vec{E}_{\vec{k}}^{(-)s}$ taken at the time moment $t$ and integrated over the time of particle motion.

As it follows from (\ref{1.2}) and  (\ref{1.3}), the expression
for the electromagnetic wave emitted by a charged particle
traveling through a crystal (natural or photonic) can be written
in the form \cite{time2}:
\begin{equation}
\label{1.7} \vec{E}(\vec{r},t)=\frac{1}{2\pi r}\sum_s \vec e^s\int
A_{\vec{k}}^s(\omega)e^{-i\omega(t-r/c)}d\omega,
\end{equation}
i.e.,
\begin{equation}
\label{1.7ad} \vec{E}(\vec{r},t)=\frac{1}{r}\sum\limits_s \vec
{e}^{~s} A^s_{\vec{k}}(t-r/c),
\end{equation}
where
\begin{equation} \label{1.7+}
A_{\vec k}^s(t-r/c)=\frac{1}{2\pi}\int A_{\vec
k}^s(\omega)e^{-i\omega(t-r/c)}d\omega, ~~ t>\frac{r}{c}.
\end{equation}

Hence, the expression for the intensity $\frac{dI(t)}{d \Omega}$
of radiation in the element of solid angle can be written in the
form
\begin{equation}
\label{1.7ad+}
 \frac{dI(t)}{d \Omega}=\frac{c}{4\pi}|\sum\limits_s \vec e^{~s}
A^s_{\vec{k}}(t-r/c)|^2.
\end{equation}

The intensity of radiation for photons having the polarization  $\vec e^{~s}$
is defined as
\begin{equation}
\label{transition0} \frac{dI_s(t)}{d \Omega}=
\frac{c}{4\pi}|A^s_{\vec{k}}(t-r/c)|^2.
\end{equation}

Equations  (\ref{1.7ad+}) and (\ref{transition0}) describe the
intensity of the pulse of radiation generated by a relativistic
particle moving along an arbitrary trajectory  $\vec{r}(t)$ in a
photonic crystal as a function of time.

%

It follows from (\ref{1.7+}) and (\ref{1.7ad+}) that  time
dependence of the  intensity $\frac{dI(t)}{d \Omega}$ of radiation
produced by a particle passing through a natural or photonic
crystal is determined by the radiation amplitude
$A_{\vec{k}}^s(\omega)$ dependence on frequency.
These amplitudes were obtained in
\cite{VG+Dub1977,VG+Grub+Dub:Ph.Lett,VG+Dub:J.Phys.C,VG+Dub1991,PXRbook,VG1997}
 for both X-ray parametric (quasi-Cherenkov) radiation and
diffraction X-ray radiation of channeled particles in crystals.


Let us consider expression
 (\ref{1.61}) for the amplitude
$A_{\vec{k}}^s(\omega)$.
To find this amplitude, we need to know the expression for the field
$\vec{E}^{(-)s*}_{\vec{k}}(\vec{r})$ at point  $\vec{r}(t)$.
The explicit expressions for $\vec{E}^{(-)s*}_{\vec{k}}$ are given
in \cite{VG_NO2012,PXRbook,VG1997}. The general expression for
$\vec{E}^{(-)s*}_{\vec{k}}$ in the crystal can be written in the
form:
\begin{equation}
\label{para_1.18}
\vec{E}^{(-)s}_{\vec{k}}(\vec{r})=\sum\limits^2_{\mu=1}\left[\vec{e}^s
\Phi_{\mu}^s e^{i\vec{k}_{\mu s}\vec{r}}+ \vec{e}^s_{\tau}
\Phi_{\tau\mu}^s e^{i\vec{k}_{\mu s\tau}\vec{r}} \right].
\end{equation}

The particle coordinate $\vec{r}(t)$ at time moment  $t$ can be written as
\begin{equation}
\vec{r}(t)=\vec{r}_0 + \vec{u} t + \delta \vec{r}(t).
\end{equation}
Here  $\vec{r}_0$ is the coordinate of the electron at time $t=0$,
$\vec{u}$ is the velocity of the electron entering the interaction
area,  and  $\delta \vec{r}(t)$ is the change in the particle
trajectory caused by the forces acting on the particle in the
interaction region.

\section{Diffraction radiation from an oscillator when particles move in a magnetostatic undulator
inside  a photonic crystal}

Let us assume for concreteness that a particle is moving in the
undulator formed by a periodic magnetostatic field or in the field
of an electromagnetic wave propagating along the direction of
particle motion that is determined by  the initial velocity
$\vec{u}$ of the particle.
Let us choose the direction of the velocity $\vec{u}$ as the $z$-axis.
The particle oscillation frequency $\Omega$ in the laboratory
frame can be written as
\begin{equation}
\Omega = \vec{\kappa} \vec{u} - \Omega_0 = \kappa_z u - \Omega_0,
\end{equation}
where $\kappa_z=\pm \kappa$ is the wave number of the
electromagnetic wave, the "+"  and  "-" signs correspond to the
cases when  $\vec{u}$ and $\vec{\kappa}$ are parallel and
antiparallel, respectively, $\Omega_0$ is the wave frequency, and
$\kappa= \frac{\Omega_0}{c} n(\Omega_0)$, where  $n(\Omega_0)$ is
the refractive index of the medium at the frequency
 $\Omega_0$.
For a static undulator, $\Omega=0$, $\kappa_z= \kappa = 2 \pi /
l_0$, where $l_0$ is the undulator period.

Thus, when a charged particle moves through the region occupied by a variable external field,
a moving oscillator having the frequency  $\Omega$ is formed.
It is natural that such an oscillator emits electromagnetic waves, whose frequency $\omega$
is determined by the Doppler effect.
\begin{equation}
\omega = \frac{\Omega}{1-\beta n(\omega) \cos \theta},
\end{equation}
where $n(\omega)$ is the refractive index of the medium and  $\theta$ is the angle between the
particle velocity  $\vec{u}$ and the direction of motion of the emitted photons.

We shall further assume that the external field (the undulator or the electromagnetic wave) leading
to particle oscillations is concentrated in the region occupied by a photonic crystal.
For concreteness, we shall further study the features of DRO
generated when a particle moves in a magnetostatic undulator
placed inside or outside a photonic crystal.

\begin{figure}[htbp]
\epsfxsize = 8 cm \centerline{\epsfbox{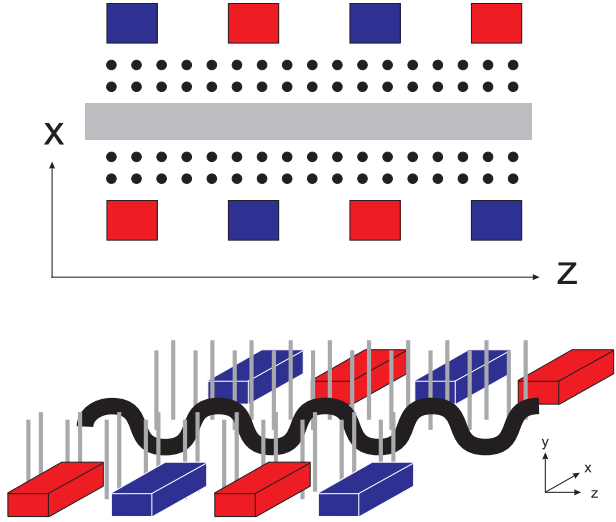}}
\caption{Undulator and photonic crystal} \label{fig:undulator}
\end{figure}

Let us suppose that a particle moves inside a magnetostatic
linearly polarized undulator  (see Fig.\ref{fig:undulator}).

According to \cite{Marshall}, the periodic magnetostatic field of the undulator in this case can approximately be  written as
\begin{equation}
\vec{B}_{\perp}(z)=\sqrt{2} B_{\perp} \vec{n}_y \sin \kappa z,
\label{eq:B-field}
\end{equation}
where $\vec{n}_y$  is the unit vector along the $y$-axis,
$B_{\perp}$ is the root-mean-square amplitude of the field 
$\kappa = \frac{2 \pi}{l_0}$, and $l_0$ is the undulator period.

As a consequence, the particle undergoes oscillations in the
$(x,z)$ plane along both the $x$ (transverse with respect to the
$z$-axis) and the $z$-axes.

Assuming that the velocity change $\delta v$, induced by the
field, is small in comparison with the initial velocity $\vec{u}$
of the particle ($\delta v \ll u$), we have for the particle
trajectory \cite{Marshall}:
\begin{equation}
\label{delta_v} x=a \sin \Omega t, ~~ z=ut-b \sin 2 \Omega t,
\end{equation}
where $a=\sqrt{2} \frac{u \Omega_{\perp}}{\gamma \Omega^2}$,
$b=\frac{u \Omega_{\perp}^2}{4 \gamma^2 \Omega^3}$, and
$\Omega_{\perp}=\frac{e B_{\perp}}{mc}$.
According to (\ref{delta_v}), the particle velocity can be
expressed as:
\begin{equation}
v_x=a \Omega \cos \Omega t, ~~ v_z=u - 2 b \Omega \cos 2 \Omega t.
\end{equation}

The field $\vec{E}_k^{(-)s*}$ in the region occupied by the
crystal (see (\ref{1.61}) and (\ref{para_1.18}))
 can be presented as %
\begin{equation}
\label{eq:para_1.18new}
\vec{E}^{(-)s*}_{\vec{k}}(\vec{r})=\sum\limits^2_{\mu=1}\left[\vec{e}^s
\Phi_{\mu}^s \exp(-i\vec{k}_{\mu s}\vec{r})+ \vec{e}^s_{\tau}
\Phi_{\tau\mu}^s \exp(-i\vec{k}_{\mu s\tau}\vec{r})\right].
\end{equation}

As a result, for the radiation amplitude we have
\begin{eqnarray}
\label{eq:**a} & &A_{\vec k}^s(\omega)= \frac{i \omega Q}{c^2} \\
& & \times \int_0^T
\left\{
\sum_{\mu=1}^2 \vec{e}^s \Phi_{\mu}^s e^{-i\vec{k}_{\mu
s}\vec{r}(t)} + \vec{e}_{\tau}^s \Phi_{\tau\mu}^s
e^{-i\vec{k}_{\mu s\tau}\vec{r}(t)}
\right\} \left( \vec{u} - \vec{n}_z 2 b \Omega \cos 2 \Omega t +
\vec{n}_x a \Omega \cos \Omega t \right) e^{i \omega t} dt,
\nonumber
\end{eqnarray}
where $\vec{r}(t)=\vec{u}t + \vec{n}_x a \sin \Omega t - \vec{n}_z
b \sin 2 \Omega t$, and
$\vec{n}_x$ and $\vec{n}_z$ are the unit vectors along the  $x$
and the  $z$ axes, respectively.

\begin{eqnarray}
\label{eq:**} A_{\vec k}^s(\omega)& = & \frac{i \omega Q}{c^2} ~
\int_0^T
\left\{
\sum_{\mu=1}^2 \vec{e}^s \Phi_{\mu}^s e^{-i\vec{k}_{\mu
s}\vec{r}(t)} + \vec{e}_{\tau}^s \Phi_{\tau\mu}^s
e^{-i\vec{k}_{\mu s\tau}\vec{r}(t)}
\right\} \vec{u}  e^{i \omega t} dt  \\
& - &  \frac{i \omega Q}{c^2}~  \int_0^T
\left\{
\sum_{\mu=1}^2 \vec{e}^s \Phi_{\mu}^s e^{-i\vec{k}_{\mu
s}\vec{r}(t)} + \vec{e}_{\tau}^s \Phi_{\tau\mu}^s
e^{-i\vec{k}_{\mu s\tau}\vec{r}(t)}
\right\}  \vec{n}_z 2 b \, \Omega \cos 2 \Omega t ~ e^{i \omega t} dt \nonumber \\
& + &   \frac{i \omega Q}{c^2}~  \int_0^T
\left\{
\sum_{\mu=1}^2 \vec{e}^s \Phi_{\mu}^s e^{-i\vec{k}_{\mu
s}\vec{r}(t)} + \vec{e}_{\tau}^s \Phi_{\tau\mu}^s
e^{-i\vec{k}_{\mu s\tau}\vec{r}(t)}
\right\} \vec{n}_x a \Omega \cos \Omega t ~ e^{i \omega t} dt ,
\nonumber
\end{eqnarray}

\noindent The exponentials in (\ref{eq:**a}) and (\ref{eq:**})
\begin{equation}
e^{-i\vec{k}_{\mu s}\vec{r}(t)}=e^{-i\vec{k}_{\mu s}\vec{u}t}
e^{-i\vec{k}_{\mu s}\vec{n}_x a \sin \Omega t}
e^{i\vec{k}_{\mu s} \vec{n}_z b \sin 2 \Omega t},
\end{equation}
\begin{equation}
e^{-i\vec{k}_{\mu s\tau}\vec{r}(t)}=e^{-i\vec{k}_{\mu s
\tau}\vec{u}t}
e^{-i\vec{k}_{\mu s \tau} \vec{n}_x a \sin \Omega t}
e^{i\vec{k}_{\mu s \tau}\vec{n}_z b \sin 2 \Omega t},
\end{equation}
can be conveniently transformed using the relation~\cite{Ryzhik}:
\begin{equation}
\label{eq:expansion}
e^{\pm i d \sin \varphi}= J_0 (d) + 2 \sum_{n=1}^{\infty} J_{2 n}
(d)~ \cos 2 n \varphi ~\pm~ 2 i \sum_{n=0}^{\infty} J_{2 n +1}
(d)~ \sin(2n+1) \varphi,
\end{equation}
or, in some cases
\begin{equation}
\label{eq:expansion1}
e^{\pm i d \sin \varphi}= e^{\pm i d \cos (\varphi+\pi/2)}=
\sum_{m=-\infty}^{\infty} (\pm i)^m J_{m} (d)~ e^{\pm i m
(\varphi+\pi/2)}.
\end{equation}

In view of (\ref{eq:expansion}) and (\ref{eq:expansion1}),  the
exponentials in (\ref{eq:**}) are expressed in terms of the series
expansions involving the product of Bessel functions into $\cos$
and $\sin$ oscillating in time at the frequencies $m \Omega$
($m=0,1,2,...$).

As a result, we have quite a cumbersome expression for  $A_k^s
(\omega)$
\begin{eqnarray}
 \label{eq:***}
& & A_{\vec k}^s (\omega) = \frac{i \omega Q}{c^2} \int_0^T
\sum_{\mu=1}^2 \vec{e}^{\, s} \Phi_{\mu}^s e^{-i \vec{k}_{\mu}^s
\vec{u} t}
\\
& & \times
\left\{ J_0 (\vec{k}_{\mu}^s \vec{n}_x a) +2 \sum_{n=1}^{\infty}
J_{2 n} (\vec{k}_{\mu}^s \vec{n}_x a) \cos 2 n \Omega t
-2 i \sum_{n=0}^{\infty} J_{2 n+1} (\vec{k}_{\mu}^s \vec{n}_x a)
\sin (2 n +1)\Omega t \right\}
\nonumber \\
& & \times
\left\{ J_0 (\vec{k}_{\mu}^s \vec{n}_z b) +2 \sum_{n=1}^{\infty}
J_{2 n} (\vec{k}_{\mu}^s \vec{n}_z b) \cos 4 n \Omega t
+2 i \sum_{n=0}^{\infty} J_{2 n+1} (\vec{k}_{\mu}^s \vec{n}_z b)
\sin 2 (2 n +1)\Omega t \right\}
\nonumber \\ & & \times \left( \vec{u} - \vec{n}_z 2 b \Omega \cos
2 \Omega t + \vec{n}_x a \Omega \cos \Omega t \right) e^{i \omega
t} dt \nonumber \\
& & \hspace{1cm} + \, \frac{i \omega Q}{c^2} \int_0^T
\sum_{\mu=1}^2 \vec{e}_{\tau}^{\, s} \Phi_{\tau \mu}^s e^{-i
\vec{k}_{\mu \tau}^s
\vec{u} t} \nonumber \\
& & \times
\left\{ J_0 (\vec{k}_{\mu \tau}^s \vec{n}_x a) +2
\sum_{n=1}^{\infty} J_{2 n} (\vec{k}_{\mu \tau}^s \vec{n}_x a)
\cos 2 n \Omega t
-2 i \sum_{n=0}^{\infty} J_{2 n+1} (\vec{k}_{\mu \tau}^s \vec{n}_x
a) \sin (2 n +1)\Omega t \right\}
\nonumber \\
& & \times
\left\{ J_0 (\vec{k}_{\mu \tau}^s \vec{n}_z b) +2
\sum_{n=1}^{\infty} J_{2 n} (\vec{k}_{\mu \tau}^s \vec{n}_z b)
\cos 4 n \Omega t
+2 i \sum_{n=0}^{\infty} J_{2 n+1} (\vec{k}_{\mu \tau}^s \vec{n}_z
b) \sin 2 (2 n +1)\Omega t \right\}
\nonumber \\ & & \times \left( \vec{u} - \vec{n}_z 2 b \Omega \cos
2 \Omega t + \vec{n}_x a \Omega \cos \Omega t \right) e^{i \omega
t} dt. \nonumber
\end{eqnarray}

%
Using the exponential representation for $\cos$ and $\sin$ and
multiplying the expressions in curly brackets one can present
(\ref{eq:***}) as sums of terms each including product of $J_n
(\vec k_{\mu}^s\vec n_z b) J_{n^{\prime}} (\vec k_{\mu}^s \vec n_x
a)$ (or $J_n (\vec k_{\mu \tau }^s\vec n_z b) J_{n^{\prime}} (\vec
k_{\mu \tau }^s \vec n_x a)$) and exponents.
%

Let us consider, for example, the terms proportional to $\vec u
J_0 (\vec k_{\mu }^s\vec n_z b) J_0 (\vec k_{\mu }^s \vec n_x a)$
and $\vec u J_0 (\vec k_{\mu \tau }^s\vec n_x a) J_0 (\vec k_{\mu
\tau }^s \vec n_z b)$ in (\ref{eq:***}) and integrate them in time
$t$. As a result, these terms will involve the factors
 $D(\omega)$, having  the form
\begin{equation}
D^s_{\mu}(\omega)=\frac{e^{-i(\vec k^s_{\mu}\vec u -
\omega)T}-1}{-i(\vec k^s_{\mu}\vec u - \omega)},\qquad
D^s_{\mu\,\tau}(\omega)=\frac{e^{-(i\vec k_{\mu\,\tau}^s\vec u -
\omega)T}-1}{-i(\vec k_{\mu\,\tau}^s\vec u - \omega)}.
\end{equation}
These factors are  sharp functions of the differences  $(\vec
k^s_{\mu}\vec u - \omega)$  and $(\vec k_{\mu\,\tau}^s\vec u -
\omega)$ and are proportional to  $T$ if
\begin{equation}
\vec k^s_{\mu} \vec u - \omega = 0 \quad\mbox{ and }\quad \vec
k_{\mu\,\tau}^s\vec u - \omega.
\end{equation}

If the conditions under which  $|\vec k^s_{\mu}| \simeq |\vec
k_{\mu\,\tau}^s|$ are fulfilled, then these conditions determine
the spectrum of parametric quasi-Cherenkov radiation (see
\cite{VG_NO2012,PXRbook}).
It is to be noted that owing to  the  transfer of energy to
harmonics, the amplitude of quasi-Cherenkov radiation, generated
by a particle moving in the undulator, is reduced by a factor of
$J_0(a)$ or $J_0(b)$ as compared to the case when the undulator is
absent.
(Compare with a similar phenomenon in the situation when a
particle, moving in the undulator, emits ordinary Cherenkov
radiation \cite{Leibowitz,Gailitis,Musikanyan,Ginzburg}).

When the absolute value of the wave vector of the diffracted
photon differs from $|\vec k^s_{\mu}|$, from the equality  $\vec
k^s_{\mu \tau} = \vec k^s _{\mu} + \vec \tau$ , we can derive the
following equation for the radiation frequency
\begin{equation}
\label{fre} \vec k^s_{\mu}\vec u -\omega = - \vec \tau \vec u
\end{equation}
that defines the spectrum of diffraction radiation (resonance and Smith-Purcell radiations).

Let a particle be incident along the normal to the surface of the
photonic crystal. In this case, we can present  (\ref{fre}) in the
form
 \begin{equation}
 \omega -  k^s_{\mu \,z} u=\tau_z,
 \end{equation}
that is,
\begin{equation}
\label{freq}
 \omega(1-\beta n\cos\vartheta) = - \tau_za,\qquad
\omega=-\frac{\tau_z}{1-\beta n(\omega) \cos\vartheta}.
\end{equation}
As is seen, in contrast to the frequency of quasi-Cherenkov radiation, this frequency increases as the particle energy increases.
%
%

Let us mention here that in considering the diffraction process
with several waves involved, the situation is possible when the
photon, emitted in the process of diffraction radiation, will
undergo diffraction from a family of planes that are defined by a
different reciprocal lattice vector $\vec{\tau}\,'$.
This case of radiation in crystals underlies the operation of the
first VFEL generator \cite{first_lasing}.

The exponentials in the remaining terms in (\ref{eq:***}) include
the power indices of the form $\vec u(\vec{k}^s_{\mu} \vec{u} -
\omega \,\pm\, m\, \Omega)t$ and $\vec u(\vec{k}_{\mu\,\tau}^{s}
\vec{u} - \omega\, \pm \,m \,\Omega)t$, which lead to the
radiation produced by the oscillator that oscillates at the
frequency $\pm \,m\,\Omega$.
%
As a result,  the  spectrum of the diffraction radiation from the
oscillator includes the harmonics whose amplitude depends on $\vec
k_{\mu}^s\vec n_x a$, $\vec k_{\mu}^s \vec n_z b$  and $\vec
k_{\mu\,\tau}^s\vec n_x a$, $\vec k_{\mu\,\tau}^s \vec n_z b$.

If the parameter $k^s_{\mu  z} b \ll 1$, i.e., the emitted photons
have a wavelength that appreciably exceeds the amplitude of
longitudinal oscillations of the particle, then the Bessel
function
 $J_0 (k^s_{\mu \,z} b) \approx 1$, while
$J_{n \ne 0} (k^s_{\mu \,z} b)\ll 1$, and the contribution to the
photon emission process from longitudinal oscillations can be
neglected.

Thus, the longitudinal oscillations of particles in undulators
that exist along with transverse ones, lead to the formation of
additional radiation peaks that are  generated through the DRO
mechanism  at large angles to the velocity of a relativistic
particle. The frequency of these peaks is larger than that
determined by the fundamental frequency $\Omega$ of oscillations.



We should  point out that, as stated earlier in this section,
owing to the transfer of energy to harmonics, the amplitude of
quasi-Cherenkov (parametric) radiation (as well as that of
diffraction radiation generated through particle motion in the
undulator) is reduced by a factor $J_0 (k^s_{\mu z} b), J_0
(k^s_{\mu x} a)$, which has an effect on the increment of
radiative quasi-Cherenkov (diffraction) instability of a beam
moving in a photonic crystal.

Analyzing  the  time evolution of radiation from a photonic
crystal, one can reveal some interesting features of radiation
produced by a particle moving in the undulator inside a photonic
crystal.
As we showed earlier
%
%
\cite{time2,time1}, in the case of quasi-Cherenkov radiation, the
law of
 radiation has a form of oscillations, whose frequency is determined by the characteristics of the crystal.
When an oscillator moves in a photonic crystal,  time oscillations
at the frequencies determined by the  frequency differences of the
harmonics $n \Omega$, where $n=1,2,3,...$ are added to those
mentioned above.

When the radiation  intensity is averaged over the time interval
being much greater than the oscillation period, these time
oscillations disappear, and only the oscillations with a longer
period remain, which  are determined by  the dispersion of the
photons diffracted in the crystal.

Now, let us give a more detailed consideration of the
time-dependent radiation amplitude $A_{\vec k}^s (t)$:
\begin{equation}
A_{\vec k}^s (t) = \frac{1}{2 \pi} \int A_{\vec k}^s (\omega)
e^{-i \omega t} d \omega \label{eq:****1}
\end{equation}

Substituting (\ref{eq:***}) into (\ref{eq:****1}) and making
integration over time as described hereinbefore one can present
the amplitude $A_k^s (t)$ in the form:
\begin{eqnarray}
& & A_{\vec k}^s (t)= \frac{1}{2 \pi} \int \sum_{\mu n n^{\prime}}
B_{\mu}^s (\omega) D_{\mu}^s (\omega,\,n,\,n') e^{-i \omega t} d
\omega \\
& & + \frac{1}{2 \pi} \int \sum_{\mu n n^{\prime}} B_{\mu \tau}^s
(\omega) D_{\mu \tau}^s (\omega,\,n,\,n') e^{-i \omega t} d
\omega, \nonumber
\end{eqnarray}
where $B_{\mu}^s (\omega)= \Phi_{\mu}^s (\omega)$ and  $B_{\mu \tau}^s
(\omega)= \Phi_{\mu \tau}^s (\omega)$ have the meaning of the coefficients of transmission and reflection from the crystal, respectively.

The quantities  $D_{\mu}^s (\omega)$ have a general form
\begin{eqnarray}
D_{\mu}^s (\omega,\,n,\, n') = \frac{e^{-i(\vec{k}^s_{\mu}
\vec{u}- \omega \pm (n+n^{\prime}) \Omega) T} - 1}{-i(\vec{k}^s_{\mu} \vec{u}- \omega \pm (n+n^{\prime}) \Omega)},\\
D^s_{\mu\,\tau}(\omega,\,n,\,n')=\frac{e^{-i(\vec
k_{\mu\,\tau}^s\vec u -\omega \pm (n+n')\Omega)T}-1}{-i(\vec
k_{\mu\,\tau}^s\vec u-\omega\pm(n+n')\Omega)}.
\end{eqnarray}

Thus, the amplitude  $A_{\vec k}^s (t)$ can be written in the
form:

\begin{equation}
A_{\vec k}^s (t)= \sum_{nn'} \int B_{\mu}^s (t-t^{\prime})
D_{\mu}^s (t^{\prime},n,n')\, d t^{\prime}
 + \sum_{nn'} \int
B_{\mu\,\tau}^s (t-t^{\prime}) D_{\mu \,\tau}^s
(t^{\prime},n,n')\, d t^{\prime},
\end{equation}
where $B(t)$ and $D(t)$ are the Fourier transforms of $B(\omega)$
and $D(\omega)$.

 Let us compare this expression with that describing the time dependence of the electromagnetic
 {pulse} $E_0 (t)$ reflected from the crystal
\begin{equation}
E (t)= \frac{1}{2 \pi} \int B_{\mu}^s (\omega) E_0 (\omega) e^{-i
\omega t} d \omega= \int B_{\mu}^s (t-t^{\prime}) E_0 (t^{\prime})
d t^{\prime}
\end{equation}
where
$E_0 (t^{\prime})= \frac{1}{2 \pi} \int  E_0 (\omega) e^{-i \omega
t^{\prime}} d \omega$, $B_{\mu}^s (\omega)$ is the coefficient of
reflection (transmission)  from the crystal of the wave  having
the frequency $\omega$.

As we can see, when radiation is excited by the particle traveling
through the crystal, the radiation pulse demonstrates similar
pattern of time dependence as that of the reflected from the
crystal pulse produced by the incident packet of electromagnetic
waves.

This might have been expected, remembering that the emission of
electromagnetic waves from a charged particle can be considered as
the diffraction in the crystal of the pseudo-photons corresponding
to this particle.
In this case the coefficient $B(t)$ describes the crystal response
function to the $\delta$-type pulse (i.e., if  $E_0 (t^{\prime})=
\delta (t^{\prime})$, then the field $E(t) = B(t)$).
 The explicit expression for the function $B(t)$ was obtained in
 \cite{time1,time2}.

Thus, the diffraction of photons emitted from a relativistic
particle moving in the undulator inside a photonic crystal leads
to complex time oscillations of the radiation amplitude.
Consequently, the radiation intensity that is proportional to
$|A(t)|^2$. In this case, the  oscillations due to the
interference between quasi-Cherenkov and undulator radiations
occur along with those due to the differences of harmonics
$n\Omega$.

Let us note here that the  amplitudes $A_k^s(\omega)$ and
$A_k^s(t)$ of radiation from a particle moving in the undulator in
the presence of a photonic crystal are determined by the
quantities $D(\omega)$ and $D(t)$ that are analogous to the
similar amplitudes of parametric X-ray radiation and X-ray
diffraction radiation from a relativistic oscillator in natural
crystals.
 For this reason, we can use the general results obtained for
 X-ray radiation and  derive, in particular, the formulas describing
 spectral-angular distribution of the radiation energy  (the
 number of emitted quanta) by
 replacing in  the formulas given in \cite{nim06} the quantities $\chi_{\tau}$, determining the
 dielectric permittivity of a natural crystal in the X-ray range,
 by the quantities $\chi_{\tau}$, determining the process of
 dynamical diffraction in a photonic crystal.
In \cite{nim06}, these quantities are given explicitly for a
photonic crystal built from metallic threads. According to the
analysis in \cite{nim06}, $\chi_{\tau}$ in such photonic crystals
depends dramatically on the polarization of the wave's electric
vector: for the waves whose electric vector $\vec E$ is parallel
to the thread, $|\chi_{\tau\parallel}| \ll |\chi_{\tau\perp}|$,
where $\chi_{\tau\perp}$ corresponds to the wave whose electric
vector is orthogonal to the thread's axis. As a consequence, when
emitted at small angles with respect to particle velocity,
quasi-Cherenkov radiation has the polarization $\vec E$ parallel
to the threads. What is more, the angular distribution  of
radiation becomes anisotropic: the radiation intensity  is sharply
suppressed if a plane is formed by vectors $\vec k$ and $\vec v$,
orthogonal to the direction of the thread, and reaches its maximum
value if this plane is parallel to the thread. This is in sharp
contrast to the angular distribution of PXR emitted at small
angles with respect to particle velocity, in which case such
anisotropy is not observed.

Let us also mention that the intensity of diffraction radiation
from a relativistic oscillator demonstrates quite  a different
time-dependence pattern than the intensity of quasi-Cherenkov
radiation, because by selecting the oscillation frequency of the
oscillator, one can tune up the center of the pseudo-photon wave
packet to such range of frequencies in which the group velocity of
the wave packet will reduce appreciably, and the radiation from
the crystal will last longer.

\section{Radiation from a dynamical undulator}

Let us note that above we have given a detailed consideration of
{particle} motion in a magnetostatic undulator, by the particle
motion in a dynamical undulator (electromagnetic wave) has some
distinct features. The trajectory of the particle moving in a
dynamical undulator includes the initial phase of wave
oscillation.  For example, the particle trajectory that describes
one-dimensional oscillations of the particle in the plane
transverse to the direction of motion, defined by the velocity
$\vec u$, can be written as

\begin{equation}
\delta\vec r_{\perp}(t)=\vec a\cos(\Omega_0 t+\delta),
\end{equation}
where $\vec a$ is the amplitude of particle oscillation in the
transverse plane and $\delta$ is the initial  oscillation phase of
the particle.
The same phase $\delta$ appears in longitudinal
oscillations of the particle.
Consequently, seeking the radiation
intensity in the case when the particle beam entering the
undulator is not modulated, we must average the the intensity over
the random phase $\delta$, and the interference between
quasi-Cherenkov and oscillator radiation (as well as the
interference of different harmonics) disappear.

In what follows, we shall consider  particle radiation in a
dynamical undulator the case of the contribution of longitudinal
oscillation of the particle can be neglected.

Let us choose $\vec u$ as the $z$-axis. Then we have for the
transverse velocity

\begin{equation}
\vec v_{\perp}= -\vec a\Omega'\sin(\Omega' t + \delta).
\end{equation}
The radiation amplitude  $A^s_{\vec k}(\omega)$ (see (\ref{1.61}))
can now be written in the form
\begin{equation}
A^s_{\vec k}(\omega)=\frac{i\omega Q}{c^2}\int\vec E_{\vec
k}^{(-)s^*}(\vec r(t)\,,\omega)\left(\vec u-\vec
a\,\Omega'\sin(\Omega' t+\delta)\right)e^{i\omega t}\,d t,
\end{equation}
i.e., in this case, the amplitude of photon emission by a charged
particle contains the contributions of two types: from parametric
quasi-Cherenkov radiation and from the radiation produced by
transverse oscillation  of a particle in the undulator.

Let us state here that a particle moving  in a natural crystal can
undergo oscillations due to the fact that it moves in a channeling
regime in a straight or a periodically bent channel (crystal
undulator)
\cite{VG+Dub1976,VG+Dub1977,Kumahov,VG+Grub+Dub:Ph.Lett,Vorobiev,Maisheev1,Maisheev2,VG+Tikh2013}.
Irrespective of what the mechanism leading to the formation of the
undulator may be, the general form of the equality (\ref{1.61})
will remain the same. Only specific characteristics of the
radiation produced by the particle moving through the crystal will
vary.

Thus, the time-dependent amplitude of radiation  $A_{\vec
k}^s(t-\frac{r}{c})$ can be presented in the form
($\tau=t-\frac{r}{c}$):
\begin{equation}
A_{\vec k}^s(\tau)=A^s_{\vec k\, par}(\tau)+A^s_{\vec k\, osc}
(\tau),
\end{equation}
where the amplitude of parametric (quasi-Cherenkov) radiation
\begin{eqnarray}
A^s_{\vec k\, \text{par}}(\tau) & = & \frac{1}{2\pi}\int A^s_{\vec k\, par}(\omega)e^{-i\omega\tau}d\omega,\nonumber\\
A^s_{\vec k\, \text{par}}(\omega) & = &
\frac{i\omega\,Q}{c^2}\int\vec E_{\vec k}^{(-)s^*}\left(\vec
r(t),\omega\right)\vec u\,e^{i\omega t} \,dt,
\end{eqnarray}
and the amplitude of diffraction radiation from a relativistic
oscillator
\begin{eqnarray}
\label{osc}
 A^s_{\vec k\, \text{osc}} (\tau)& =
&\frac{1}{2\pi}\int A^s_{\vec k\, osc} (\omega)
e^{-i\omega\tau} d\omega,\nonumber\\
A^s_{\vec k\, \text{osc}} (\omega) & = &\frac{i\omega\,Q}{c^2}\int
\vec E_{\vec k}^{(-)^*}\left(\vec r(t)\,,\omega\right)\left(-\vec
u\,\Omega'\sin(\Omega't+\delta)\right) e^{i\,\omega t} dt.
\end{eqnarray}
The time-dependent behavior of parametric quasi-Cherenkov
radiation have been studied in the previous section. Let us give
now  a more detailed consideration of diffraction radiation from
an oscillator.


For example, it follows from \cite{time2} that in the Laue case,
the expression $\vec E_{\vec k}^{(-)s^*}$ appearing in the
equation for the radiation amplitude (\ref{osc}) has the following
form inside the crystal
\begin{eqnarray}
\vec E_{\vec k}^{(-)s^*}(\vec r)=\vec
e\,^s\left[-\sum_{\mu=1}^2\xi^0_{\mu\,s}e^{i\frac{\omega}{\gamma_0}\varepsilon_{\mu
s}(L-z)}\right]e^{-i\vec k\vec r}\nonumber\\ +\vec
e\,^s_{\tau}\beta_1 \times\left[\sum_{\mu=1}^2\xi_{\mu
s}^{\tau}e^{i\frac{\omega}{\gamma_0}\varepsilon_{\mu
s}(L-z)}\right]e^{-i\vec k_{\tau}\vec r},
\end{eqnarray}
%
%
where
\[
\vec{k}_{\mu s}=\vec{k}+{\kappa}^*_{\mu s}\vec{N},\qquad
\kappa_{\mu s}^*=\frac{\omega}{c\gamma_0}\varepsilon^*_{\mu s},
\]
$\mu=1,2$; $\vec{N}$ is the unit vector of a normal to the
entrance crystal surface which is directed into the crystal,
\[
\varepsilon_{1(2)s}=\frac{1}{4}\left[(1+\beta_1)\chi_0-\beta_1\alpha_B\right]
\pm\frac{1}{4}\left\{\left[(1-\beta_1)\chi_0+\beta_1\alpha_B\right]^2+4\beta_1C_s^2\chi_{\vec{\tau}}\chi_{\vec{-\tau}}\right\}^{-1/2},
\]
$\alpha_B=(2\vec{k}\vec{\tau}+\tau^2)k^{-2} $ is the off-Bragg
parameter ($\alpha_B=0$ if the exact Bragg condition of
diffraction is fulfilled),
\[
\gamma_0=\vec{n}_{\gamma}\cdot\vec{N},\quad
\vec{n}_{\gamma}=\frac{\vec{k}}{k},\quad
\beta_1=\frac{\gamma_0}{\gamma_1}, \quad
\gamma_1=\vec{n}_{\gamma\tau}\cdot\vec{N},\quad
\vec{n}_{\gamma\tau}=\frac{\vec{k}+\vec{\tau}}{|\vec{k}+\vec{\tau}|}.
\]

Note that the trajectory $\delta \, r_{\perp} (t)$, which depends
on the external field, enters into the exponent, and this leads to
the appearance of the exponentials of the form
$e^{i\,b\cos\varphi}$. For further analysis, it is convenient to
use the following equality \cite{Ryzhik}
\begin{equation}
e^{i\,b\cos\varphi}=\sum_{m=-\infty}^{\infty}i^m
J_m(b)e^{im\varphi}.
\end{equation}
We can write as a result
\begin{eqnarray}
A^s_{\vec k}(\omega) & =
&\frac{i\omega\,Q}{c^2}\int\limits^{T}_{0}\left(\vec u-\vec
a\Omega'\sin(\Omega' t+\delta)\right)\vec E_{\vec
k}^{(-)^*}\left(\vec r(t),\, \omega\right) e^{i\omega
t}\,dt\nonumber\\
& = & \frac{i\omega\,Q}{c^2}\int\limits^T_0\left(\vec u-\vec
a\Omega'\sin(\Omega' t+\delta)\right)\nonumber\\
& \times & \left\{\vec e\,^s\left[-\sum^2_{\mu=1}\xi_{\mu
s}^{0}e^{i\frac{\omega}{\gamma_0}\varepsilon_{\mu s}(L-u
t)}\right]
e^{-i(\vec k(\vec u t+\vec a\cos(\Omega't+\delta)))}\right.\nonumber\\
& + & \left.\vec e\,^s_{\tau} \beta_1\left[\sum_{\mu=1}^2\xi_{\mu
s}^{\tau}e^{i\frac{\omega}{\gamma_0}\varepsilon_{\mu
s}(L-ut)}\right]e^{-i\vec k_{\tau}(\vec u t+\vec a\cos(\Omega'
t+\delta))}\right\}e^{i\omega t} dt.
\end{eqnarray}
Make use of the relationships
\begin{equation}
e^{ib\cos\varphi}=\sum^{+\infty}_{m=-\infty}i^m J_m(b)e^{i
m\varphi},\qquad
e^{-ib\cos\varphi}=\sum^{+\infty}_{m=-\infty}(-i)^m
J_m(b)e^{-im\varphi}.
\end{equation}
We have
\begin{eqnarray}
A^s_{\vec k}(\omega) &= &
\frac{i\omega\,Q}{c^2}\int\limits_0^T(\vec u-\vec
a\Omega'\sin(\Omega' t+\delta))\\
& \times & \left\{\sum^{\infty}_{m=-\infty}(-i)^m J_m(\vec k\vec
a)\vec e\, ^s\left[-\sum^2_{\mu=1}\xi_{\mu
s}^{0}e^{i\frac{\omega}{\gamma_0}\varepsilon_{\mu s}(L-u
t)}\right] e^{-i\vec k\vec u t} e ^{-im\Omega' t}
e^{im\delta}\right.\nonumber\\
& + & \left.\sum^{+\infty}_{m=-\infty}(-i)^m\, J_m(\vec
k_{\tau}\vec a)\vec e\,^s_{\tau}\beta_1
\left[\sum^2_{\mu=1}\xi_{\mu s}
^{\tau}e^{i\frac{\omega}{\gamma_0}\varepsilon_{\mu s}(L-u
t)}\right] e^{-i\vec k_{\tau}\vec u t} e^{-im\Omega' t} e^{-i
m\delta}\right\}e^{i\omega t} dt\nonumber.
\end{eqnarray}
These expressions include the time-dependent exponential
$e^{-i\frac{\omega}{\gamma_0}\varepsilon_{\mu\,s}ut}e^{-i\vec
k\vec u t}e^{-im\Omega' t}$ (in the second term $\vec k_{\tau}\vec
u$ appears instead of $\vec k u$ ).  Let us note here that
 $\sin(\Omega' t+\delta)=\frac{e^{i(\Omega'
t+\delta)}-e^{-i(\Omega' t+\delta)}}{2i}$. As a result, $(\vec
u-a\Omega'\sin(\Omega' t+\delta))$ is divided into three terms
$\left(\vec u-a \Omega'\frac{e^{i(\Omega' t+\delta)}-e^{-i(\Omega'
t+\delta)}}{2i}\right)$ which are multiplied by a figure bracket
$\{\ldots\}$ and  $e^{i\omega t}$.

Let also take into account that  $\vec
k+\frac{\omega}{\gamma_0}\varepsilon_{\mu \,s}=\vec k_{\mu\, s}$.

Let us first consider radiation in the forward direction. The
radiation in the direction of diffraction can be obtained through
the replacement
 \[
\vec e_s\rightarrow \beta_1\vec e\,^s_{\tau},\qquad \vec
k\rightarrow\vec k_{\tau}\qquad \mbox{и}\qquad
\xi^0_{\mu\,s}\rightarrow \xi^{\tau}_{\mu\,s}.
\]
Now we have for the amplitude in forward direction
\begin{eqnarray}
 & & A_{\vec k \, fw}^s(\omega)
  =
\frac{i\omega\,Q}{c^2}\int\limits^T_0\left(\vec u-\frac{\vec
a\Omega'}{2i}e^{i(\Omega' t+\delta)}+\frac{\vec
a\Omega '}{2i}e^{-i(\Omega' t+\delta)}\right)\nonumber\\
 & & \times
 \sum_{m=-\infty}^{\infty}(-i)^m J_m(\vec k\vec a)\vec
e\,^s\left[-\sum^2_{\mu=1}\xi^{0}_{\mu
s}e^{i\frac{\omega}{\gamma_0}\varepsilon_{\mu\, s}L}e^{-i(\vec
k_{\mu s}\vec
u+m\Omega'-\omega)t}\right]e^{-im\delta}dt \\
&  & =
  \frac{i\omega\,Q}{c^2}\left\{\sum^{\infty}_{m=-\infty}(-i)^m
J_m(\vec k\vec a)(\vec u\vec e\,^s)\left[-\sum_{\mu=1}^2\xi_{\mu
s}^{0}e^{i\frac{\omega}{\gamma_0}\varepsilon_{\mu\,
s}L}\,\frac{e^{i\Delta^m_{\mu \,s}
(\omega)T}-1}{i\Delta^m_{\mu\,s}(\omega)}\right]
e^{-im\delta}\right.\nonumber\\
& & -
 \sum^{\infty}_{m=-\infty}(-i)^m J_m(\vec k\vec
a)\left(\frac{(\vec a\vec
e\,^s)\Omega'}{2i}\right)\int\limits^T_0\left[-\sum_{\mu=1}^2\xi_{\mu\,s}^{0}\,e^{i\frac{\omega}{\gamma_0}\varepsilon_{\mu
s}L}e^{-i(\vec k_{\mu \,s}\vec
u+(m-1)\Omega'-\omega)t}\right]e^{-i(m-1)\delta}dt\nonumber\\
& & +
 \left.\frac{\vec a\vec
e\,^s\Omega'}{2i}\sum^{+\infty}_{m=-\infty}(-i)^m J_m(\vec k\vec
a)\left[-\sum_{\mu=1}^2\xi_{\mu\,
s}^{0}e^{i\frac{\omega}{\gamma_0}\varepsilon_{\mu\,s}L}e^{-i(\vec
k_{\mu\,s}\vec u+(m+1)\omega'-\omega
t)}\right]e^{-i(m+1)\delta}\,dt\right\}\nonumber
\end{eqnarray}
where
\[
\Delta^m_{\mu\,s} (\omega)=\omega-\vec k_{\mu \,s}\vec u- m\Omega'
\]
Finally,
\begin{eqnarray}
& & A_{\vec k \, fw}^s(\omega)  = \frac{i\omega\,Q}{c^2}
\sum_{m=-\infty}^{\infty}(-i)^m J_m (\vec k \vec a)(\vec u \vec
e\,^s)\nonumber\\
&& \times \left[-\sum^2_{\mu=1}\xi^{0}_{\mu
s}e^{i\frac{\omega}{\gamma_0}\varepsilon_{\mu\,
s}L}\frac{e^{i\Delta^m_{\mu\,s}(\omega)T}-1}{i\Delta^m_{\mu\,s}(\omega)}\right]
e^{-i
m\delta}\\
& &- \frac{i\omega\,Q}{c^2}\sum^{\infty}_{m=-\infty}(-i)^m
J_m(\vec k\vec a)\left(\frac{\vec a\vec e \,^s\Omega'}{2i}\right)
\left[-\sum_{\mu=1}^2\xi_{\mu
s}^{0}e^{i\frac{\omega}{\gamma_0}\varepsilon_{\mu\,
s}L}\,\frac{e^{i\Delta^{m-1}_{\mu \,s}
(\omega)T}-1}{i\Delta^{m-1}_{\mu\,s}(\omega)}\right]
e^{-i(m-1)\delta}\nonumber\\
& & +  \frac{i\omega\,Q}{c^2}\sum^{\infty}_{m=-\infty}(-i)^m
J_m(\vec k\vec a)\left(\frac{\vec a\vec e \,^s\Omega'}{2i}\right)
\left[-\sum_{\mu=1}^2\xi_{\mu
s}^{0}e^{i\frac{\omega}{\gamma_0}\varepsilon_{\mu\, s} L}\,
\frac{e^{i\Delta^{m+1}_{\mu \,s}
(\omega)T}-1}{i\Delta^{m+1}_{\mu\,s}(\omega)}\right]
e^{-i(m+1)\delta}\nonumber.
\end{eqnarray}

The amplitude  $A_{\vec k \, dif}^s(\omega)$ is obtained by
replacing $\vec e_s \rightarrow \beta_1 \vec e\,^s_{\tau}$ and
$\vec k \rightarrow \vec k_{\tau}$, $\xi^0_{\mu s}\rightarrow
\xi^{\tau}_{\mu s}$.

Now let us point out that in the first sum, proportional to $(\vec
u\,\vec e\,^s)$, the term  that includes  $m=0$ describes
 parametric quasi-Cherenkov radiation in the undulator. The
 amplitude of this radiation can be described as
$A_{par}(\omega) \cdot J_0(\vec k\vec a)$, where $A_{par}(\omega)$
is the amplitude of parametric quasi-Cherenkov radiation of the
particle moving with constant velocity $\vec u$
\cite{PXRbook,VG_NO2012}.

In the second sum, the term including  $m=1$ coincides with the
amplitude of ordinary parametric radiation if we replace $(\vec u
\vec e\,^s)$ by $i J_1(\vec k\vec a)\frac {1}{2i}(\vec a\vec
e\,^s)\Omega'$.

In the third sum, the term including $m=-1$ coincides with the
amplitude  parametric radiation if we replace $(\vec u \vec
e\,^s)$ by $i J_{-1}(\vec k \vec a)\frac {1}{2i}(\vec a\vec
e\,^s)\Omega'$.

Thus, the general expression for the amplitude of parametric
radiation can be obtained from that for the ordinary amplitude by
replacing $(\vec u \vec e\,^s)$ with the following sum:
\begin{equation*}
\vec u \vec e\,^s \rightarrow \vec u \vec e\,^s J_0(\vec k \vec a)
+ \frac{1}{2} J_1 (\vec k \vec a)\Omega' (\vec a\vec e\,^s) +
\frac{1}{2} J_{-1} (\vec k \vec a)\Omega' (\vec a\vec e\,^s)
\end{equation*}
But the sum of the last two terms is zero, since $J_{-1} = -J_1$.

So in the second and third sums, we need to remove the terms with
$m=1$ (in the second) and $m=-1$ (in the third), because their sum
equals zero.

Let us note now that in order to obtain the radiation intensity,
we need to  square the absolute value of  the amplitude $A_{\vec
k}^s$.
As a result, we have double sums over $\sum_m\, \sum_{m'}$, which
will initiate oscillations determined by the interference of
different harmonics  and the interference between quasi-Cherenkov
and undulator radiations.
The conclusion about the possibility to observe these oscillations
depends on whether the phase $\delta$ is random or not. The phase
$\delta$ will be  random unless special measures are taken, and
averaging over $\delta$  will then lead to the absence of the
above interference. So it will suffice to study the time
dependence of the radiation that is due to a certain harmonic $m$.
The intensity is then equal to  the sum of time-dependent
intensities of quasi-Cherenkov radiation and diffraction radiation
from a relativistic oscillator formed in a dynamical undulator
(diffraction radiation of a channeled particle or a particle
moving in a crystal undulator).

It is interesting to note that the spectrum of radiation from
relativistic protons or nuclei channeled in a crystal lies in the
soft spectral range, where the diffraction of the emitted photons
can be carried out in a large number of  various photonic
crystals. This enables one to use such crystals for a thorough
study of quasi-Cherenkov (parametric) radiation and diffraction
radiation from relativistic oscillators formed by channeled
protons or nuclei in both frequency and time domains.


\section{Conclusion}

The radiation amplitudes have been obtained that determine
spectral-angular and time features of radiation from a
relativistic particle moving in the undulator placed in (or near)
a photonic crystal.
Worthy of mention is that for photonic crystals built from
metallic threads, the dynamical diffraction of electromagnetic
waves has appreciable effects even in the diffraction cases that
correspond to large values of $\vec{\tau}$, i.e., for photons with
the wavelength $\lambda \ll d$ ($d$ is the lattice period; for a
rectangular lattice $\vec \tau = (\frac{2\pi n}{d}, \,\frac{2\pi
n_2}{d_2},\, \frac{2\pi n_3}{d_3})$).
As a result,the effective diffraction reflection   in such
structures (see Fig.{\ref{fig:undulator}}) is  possible,
particularly in the terahertz range, even if the lattice period of
the photonic crystal is within the centimeter range.

Let an undulator generating radiation in a terahertz range have a
period of 10 cm and the length of 1 m, while a photonic crystal
have the same length and a period of 1 cm.
Then for the reciprocal lattice vector $\vec\tau$ with $n = 30$,
we have the diffraction reflection in the terahertz range, and so
all the phenomena that are due to dynamical diffraction can be
observed in this frequency range without the necessity to use a
photonic crystal having a period within a submillimeter range.
An appreciably simpler photonic crystal
is quite adequate for such observations.
 Due to
greater values of the increment of radiative instability, volume
free electron lasers designed on the basis of this mechanism will
have much smaller undulators and  photonic crystals than those in
FELs without such crystals.
 Let us point out in this connection that multi-wave cases of diffraction of radiation in photonic crystals have a
 contribute significantly to the reduction of the resonator size and the generation threshold \cite{review}.
This circumstance can also substantially  simplify the design and
development of a two-stage VFEL for a short-wave range
\cite{two-stageVFEL}.

\end{document}